# Mechanism of hydrogen adsorption on gold nanoparticles and charge transfer probed by anisotropic surface plasmon resonance

William Watkins and Yves Borensztein

*Sorbonne Universités, UPMC Univ Paris 06, CNRS-UMR 7588, Institut des NanoSciences de Paris, F-75005, Paris, France.*

william.watkins@insp.upmc.fr ; yves.borensztein@insp.upmc.fr

The adsorption of hydrogen on Au nanoparticles (NPs) of size of the order of 10 nm has been investigated by use of localised surface plasmon resonances (LSPR) in the NPs. The samples, formed by Au NPs obtained by oblique angle deposition on glass substrates, display a strong optical dichroism due to two different plasmon resonances dependent on the polarisation of light. This ensured the use of Transmittance Anisotropy Spectroscopy, a sensitive derivative optical technique, which permitted one to measure shifts of the LSPR as small as 0.02 nm upon H adsorption, which is not accessible by conventional plasmonic methods. The measured signal is proportional to the area of the NPs, which shows that H atoms diffuse on their facets. A negative charge transfer from Au to H is clearly demonstrated.

## Introduction

Gold can be considered as the "noblest of all metals" and chemisorption of molecules, including dihydrogen, on its surface is not favoured [1,2]. Indeed, it has experimentally been shown, in agreement with theoretical results [3,4] that $H_2$ cannot chemisorb at room temperature on extended gold surfaces [5,6]. However, the situation is quite different with Au nanoparticles (NPs) of size from 2 to about 10 nm, and consequently Au NPs-based heterogeneous catalysis has become a very rich and promising field of research. Au NPs can be very reactive and can dissociate $H_2$, hence revealing a high catalytic efficiency for hydrogenation reactions [7–9]. Dissociative adsorption of $H_2$ on Au NPs deposited on different oxides has been experimentally shown by X-ray absorption spectroscopy [10], Hydrogen/Deuterium (H/D) exchange experiments [10–12] and infra-red measurements [13,14]. There is a general experimental and theoretical agreement that the active sites are the low-coordinated atoms located on the corners or edges of the Au NPs [3,4,10,15]. Nevertheless, the mechanism of $H_2$ dissociation and the fate of the H atoms are still not completely understood, which are of crucial importance for better understanding the Au-based catalytic hydrogenation reactions. For example, discrepancies exist regarding which atoms of the NPs are specifically the most reactive, whether there is a charge transfer between gold and H, whether the dissociation is spontaneous [3], almost spontaneous [4] or activated [10,12], or whether the H atoms are mobile on the Au NPs. Moreover, in spite of the numerous studies undertaken in this field, to our knowledge the issue of diffusion and recombination of the H atoms after $H_2$ dissociation has not been investigated so far. A comprehensive investigation of these latter issues is therefore still essential for a deep understanding of H adsorption and hydrogenation reactions on Au NPs. In the present article, we present a detailed investigation of the adsorption and diffusion of $H_2$ on Au NPs, by following the wavelength shift of the localised surface plasmon resonance (LSPR) displayed by the NPs, which is explained by charge transfer between Au and H.

Indeed, Au NPs display peculiar physical properties related to their small size [16–19] and specifically they can support the LSPR, giving rise to optical absorptions which can be tuned at different wavelengths as a function of their size, shape and environment [17], which can be observed with either far field or near field methods [20]. The wavelength position of the LSPR being very sensitive to the NPs' surrounding medium, biological and chemical plasmonic sensors have been developed on the basis of Au NPs [21–23]. Hence, the high sensitivity of the Au NPs LSPR upon gas adsorption makes it an efficient tool for gathering information on their chemical properties. In particular, the red or blue shift of the LSPR, and its amplitude, are directly related to the sign and to the amount of charge transfer between Au and adsorbed species, and should inform on the charge transfer between Au and adsorbed H.[24,25]

However, the amount of adsorbed H atoms on the Au NPs, hence the transferred charge, is expected to be small, and thus the shift difficult to measure. Most of LSPR-based sensors rely on the shift in wavelength of the position of the LSPR, $\Delta\lambda$, induced by the analyte adsorption. This shift is usually determined by spectroscopic measurements around the resonance, followed by a fitting of the curve with e.g. a Gaussian function. Such a method is limited by the resolution and the reproducibility of the spectrometer which are, for current monochromators, between 0.1 nm and 1 nm. A much better sensitivity can be reached by using differential techniques. In the present investigation, the strategy we have chosen was to use a highly sensitive optical method, the Transmittance Anisotropy Spectroscopy (TAS), which is capable of determining minute changes in the optical response of transparent samples, induced by the adsorption of molecules or atoms. The use of TAS on small (10 nm) anisotropic Au NPs prepared by oblique angle deposition on glass allowed us to reach an increased sensitivity on the LSPR shift and, hence, the clear observation of small hydrogen adsorption. We report here the experimental observation of shifts of the LSPR ranging from 0.02 nm to 1 nm, upon $H_2$ exposure of the Au NPs, which is explained by a negative charge transfer between Au and adsorbed $H_2$, and by diffusion of adsorbed H atoms on the Au NPs' facets.

## Principle of the method and experimental details

**TAS for investigating anisotropic gold NPs**

The TAS technique is derived from the Reflectance Anisotropy Spectroscopy (RAS), which measures the reflectance anisotropy of surfaces [26–28]. This optical method is very sensitive to the surface structure and to the adsorption of atoms or molecules on anisotropic surfaces. However, it has mainly been used on flat surfaces, either in vacuum [27,29–32], gas [33,34] or liquid [35,36], yet only few investigations focusing on the LSPR of supported metal NPs have been reported [37–41]. Recently, the LSPR of anisotropic metal NPs, synthesised by colloidal lithography and placed in a solution, have been measured using RAS [40], which showed small LSPR shifts upon changing the refraction index of the solution or upon adsorbing molecules on the NPs. However, in order to investigate the interaction of $H_2$ with Au NPs of relevant size for catalysis, it is necessary to obtain anisotropic NPs with sizes of a few nms, which is not feasible with lithographic techniques. For this reason, the choice here was to prepare small Au NPs by oblique angle deposition on a glass substrate at room temperature. Not only does this method yield anisotropic samples formed of NPs with lateral size ranging from 5 to 20 nm and height around 5 nm, but as the so-obtained samples are transparent, the use of a mirror behind the sample can be used to measure the transmittance anisotropy (TA) squared rather than the reflectance, as shown in the schematic of Fig.1, which leads to an increase of sensitivity.

**Elaboration of the anisotropic gold samples**

The Au samples were prepared by oblique angle deposition on glass substrates under vacuum of $3.10^{-6}$ mbar. Microscope slides were used as substrates. They were cleaned using ethanol and loaded into the evaporating chamber. The evaporation was carried out from a crucible heated by direct current. The sample was positioned at a 10° grazing angle to the crucible and the evaporation rate was 0.02 nm/s. The mass thickness was controlled using a calibrated quartz balance and corrected for taking into account the angle of evaporation. The characterisations of the samples were carried out by optical measurements, scanning electron microscopy (SEM) and atomic force microscopy (AFM). Several samples were elaborated with different mass thicknesses ranging from 0.5 to 3 nm.

**Scanning Electron Microscopy (SEM) and Atomic Force Microscopy (AFM)**

The SEM apparatus used was a Zeiss Supra 40. The settings were EHT = 3 and 6 kV at a working distance of 4 to 5 mm with an aperture of 7 µm. The SEM needed to be adjusted with a low potential, as the substrate being nonconductive, the samples were prone to charging. The AFM was a Digital Instruments dimension 3100 atomic force microscope, used in tapping mode.

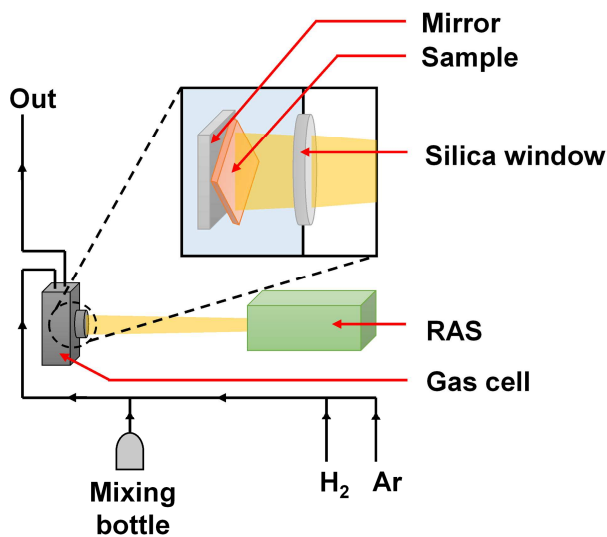

Fig.1. Schematic of the set-up used for measuring the TA spectrum of a sample under a controlled atmosphere of Ar and $H_2$. A mixing bottle is used in order to get accurate mixtures of $H_2$ in Ar. A mirror placed in the gas cell reflects the light, which leads to the measurement of the TA signal squared.

**Optical measurements**

The optical transmissions were measured for each sample on a home-made bench uv-vis spectrometer. Both parallel and perpendicular polarisations in relation to the sample's direction of evaporation were recorded. The spectrometer used was a Maya 2000 Pro from Ocean Optics Cie., lit with deuterium and quartz-halogen lamps. The anisotropic measurements were performed on a home-made RAS system, designed following the schematic described by Aspnes [26], with a 1/8 m Cornerstone130 monochromator from Newport and a 1200 lines/mm holographic grating, which nominal wavelength resolution is around 4 to 6 nm with the 0.5-1 mm slits used here. A 1 s time constant was used on the lock-in amplifier through out the experiments, which insured a time response of about 3 s.

**Gas flow system**

The samples were loaded in a gas cell equipped with a silica window enabling the TAS measurements during gas cycles. The samples were exposed at atmospheric pressure to flows of ultra-pure Ar and $H_2$ at 1000 sccm. These large flows were used to rapidly purge the lines and the gas cell. Prior to the experiments, the whole system was purged with cycles of $H_2$ and Ar for a full day in order to eliminate any gas contaminants, mainly water and oxygen. Experiments were performed for several days, during which only Ar and $H_2$ were introduced in the cell. Long cycles of $H_2$ and Ar were done till a stable behaviour was reached, after which cycles of 100 s of Ar and $H_2$ were undertaken. Accurate $H_2$ / Ar gas mixtures were prepared by using an additional bottle in order to expose the sample to different partial pressures of $H_2$ at atmospheric pressures. The cell was equipped with a heating element and a thermocouple.

# Results and discussion

**Structural and optical characterization**

Several samples with increasing amounts of Au were investigated, which qualitatively display the same behaviour, hence detailed results are only given for one sample, which was prepared with an Au mass thickness of 1.4 nm. Results obtained for the other samples are presented in Fig.S3 of the Supplementary Information. Fig.2.a shows the SEM image analysis performed on this sample. The Au NPs have a lateral size distribution ranging from 5 to 21 nm, with an average of 12 nm (Fig.2.c.). Some of the NPs display apparent irregular shapes, though, due to the charging of the sample during the SEM measurements, the resolution of the smaller NPs is limited. It is likely that these apparent irregularities are rather formed by agglomeration of a few more regular and smaller NPs (Fig.2.a). The average particle height was determined from the AFM measurements and equal to 4.5 nm, indicating that they display flat morphologies. No apparent elongated shapes with a preferential orientation can be observed with the present resolution. Nevertheless, some chains of NPs are visible, with orientations ranging between approximately +30 ° and -20° with respect to the horizontal line. Examples of such lines of NPs are indicated by the arrows on Fig.2.a. The 2D Fourier transform of this image does indeed reveal a small anisotropy (Fig.2.b). Despite this small apparent anisotropy, the optical measurements show a clear polarisation dependence. The optical square transmittance of the sample, for polarisations perpendicular ($T_\perp$) and parallel ($T_{//}$) to the direction of deposition is shown in Fig.3.a. It displays an important dichroism, with sharp LSPR given by the minimum of the spectra and centred at 600 nm for $T_\perp$ and at 700 nm for $T_{//}$. These LSPR correspond to flat Au NPs, of aspect ratio (height/lateral size) around 0.37 and a small overall anisotropy, caused by a slight anisotropic shape of the NPs and/or interaction between the NPs along the chains (see Supplementary Information).

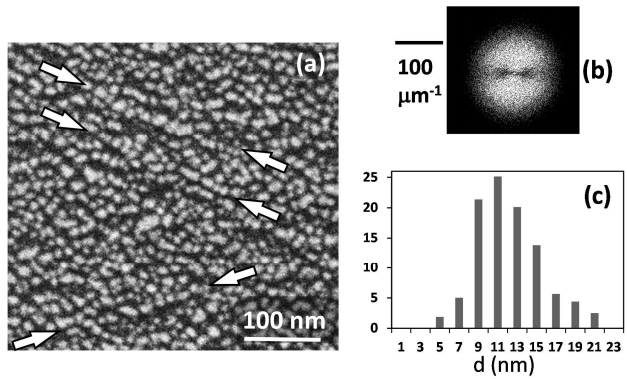

Fig.2. (a) Image obtained by SEM of Au on glass. (b) Fourier Transform of the SEM image. (c) Histogram giving the frequency as a function of the lateral size of the NPs. The sample was prepared by oblique angle evaporation of Au on glass. The direction of evaporation was aligned to the vertical axis of the image from bottom to top.

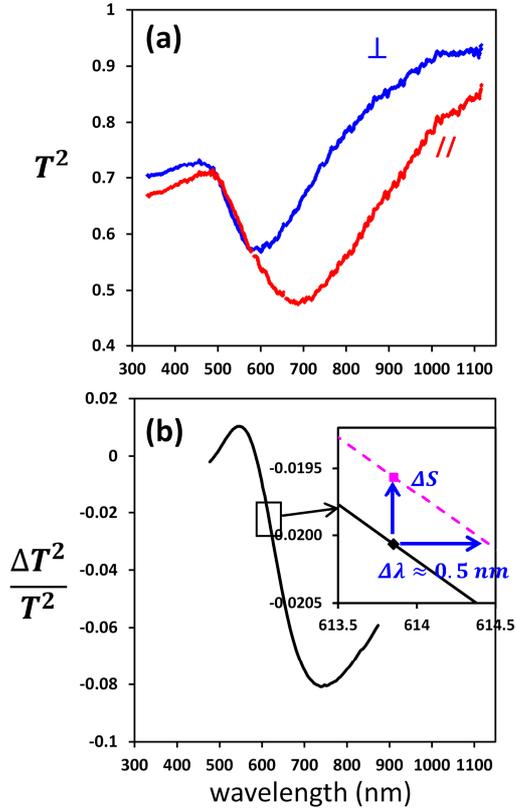

Fig.3. (a): Square of the transmittance of the sample (normalised to the glass transmittance) for polarisation of light parallel (red) and perpendicular (blue) to the direction of evaporation. (b) Square transmission anisotropy spectrum (TAS) of the sample under Ar (solid line). Insert: schematic zoom around 613 nm. A large dichroism is displayed in (a) demonstrating the anisotropy of the sample. When drawing the TA spectrum (b) under $H_2$ (dashed line), there is a clear shift of about 0.5nm, seen in the insert, when compared to the spectrum under Ar (solid line). This measured shift is smaller than the nominal resolution of the monochromator, as it is explained in the text.

**TAS measurements**

The corresponding square transmittance anisotropy spectrum, defined by

$$\frac{\Delta T^2}{T^2} = \frac{T_\perp^2 - T_\parallel^2}{1/2(T_\perp^2 + T_\parallel^2)} \qquad (5)$$

is measured directly by the TAS apparatus and is drawn in Fig.3.b., displaying a large signal/noise ratio. The TA spectrum under Ar shows a maximum at 540 nm and a minimum at 740 nm. An overall shift to longer wavelengths $\Delta\lambda$ is hardly measured at this scale upon $H_2$ exposure. This shift is due to the charge transfer between Au and H, as explained in next paragraph. Instead of measuring this small shift $\Delta\lambda$ smaller than 1 nm, which would need the use of a very high resolution monochromator, it is both more accurate and more convenient to measure the change of intensity $\Delta S$ of the TA at a fixed wavelength, e.g. around 613 nm. This is illustrated in the insert of Fig.3.b., which shows a schematic zoom of the spectrum before and after the redshift due to $H_2$ exposure. Measuring $\Delta S$ at a fixed wavelength, instead of $\Delta\lambda$, enables one to go beyond these limitations inherent to monochromators, as it is not a spectral measurement anymore. This is the key point of the present approach, which permits one to reach a very high sensitivity upon adsorption. In the present experiments, the effective wavelength resolution is as low as 0.05 nm, two orders of magnitude better than the nominal resolution of the present spectrometer. Furthermore, this differential method insures a high stability of the measurements, since it is not perturbed neither by the light's fluctuations in intensity nor by fluctuations in the optical alignment. The limitation in the accuracy is now mainly due to the sensitivity and the dynamics of the photodetector and to the quality of its electronic chain. With the present sample, the rms noise in the TA is smaller than $1.10^{-5}$, and the ultimate sensitivity reached is better than $5.10^{-5}$, which is equivalent to a wavelength shift $\Delta\lambda$ smaller than 0.05 nm, comparable to the resolution and reproducibility reached by the high end monochromators.

**Modelling the charge transfer between adsorbates and Au**

This part focuses on the effect that charge transfers between adsorbates and Au NPs have on the LSPR position. The dielectric function of Au is well described with a Drude component accounting for the s-p conduction electrons and an interband component $\varepsilon^{ib}(\omega)$ which is mainly due to the electronic transitions involving the electrons of the d bands:

$$\varepsilon(\omega) = 1 - \frac{\Omega_p^2}{\omega(\omega + i\tau^{-1})} + \varepsilon^{ib}(\omega) \qquad (1)$$

where the damping term, $\hbar\tau^{-1}$, and the plasma frequency, $\hbar\Omega_p$, can be determined from the dielectric function's experimental values obtained at room temperature [42] giving $\hbar\tau^{-1}$ = 0.072 eV and $\hbar\Omega_p$ = 8.92 eV [43]. The plasma frequency is given by:

$$\Omega_p = \left(\frac{N e^2}{\varepsilon_0 m}\right)^{\frac{1}{2}} \qquad (2)$$

where $e$ and $m$ are the charge and the mass of the electron, respectively, and N is the conduction electron density. The value of $\hbar\Omega_p$ corresponds to a density N equal to $5.78.10^{28}$ m$^{-3}$, hence the effective number of conduction electrons is equal to 0.98 per Au atom.

Considering the charge transfer from Au to an adsorbate as a variation $\Delta N$ of the conduction electron density in Au, the modified plasma frequency is obtained by replacing N by $N + \Delta N$ in Eq. (2); this modifies the dielectric function of Au and therefore induces a shift of the LSPR. In the case of an ellipsoidal Au particle with a plasmon resonance in direction j, the position $\omega_{LSPR}$ of the LSPR is given by the pole of the polarisability:

$$\alpha_j = V \frac{\epsilon(\omega) - \epsilon_m}{\epsilon_m + L_j(\epsilon(\omega) - \epsilon_m)} \qquad (3)$$

where $\epsilon_m$ is the dielectric function of the embedding medium and $L_j$ is the depolarisation factor in direction j [44]. Simple algebra shows that the relative shifts in $\omega_{LSPR}$ and in the LSPR wavelength $\lambda_{LSPR}$ are given by:

$$\frac{\Delta\lambda_{LSPR}}{\lambda_{LSPR}} = -\frac{\Delta\omega_{LSPR}}{\omega_{LSPR}} = -\frac{\Delta\Omega_p}{\Omega_p} = -\frac{1}{2}\frac{\Delta N}{N} \qquad (4)$$

As a consequence, an electron transfer from Au to the adsorbate ($\Delta N < 0$) should yield a shift of the plasma frequency and therefore a shift of the LSPR towards lower energies i.e. longer wavelengths, known as a redshift [24]

($\Delta\lambda_{LSPR} > 0$). Such shift is expected to be small. Typically, for a sphere of diameter 12 nm, the ratio between the numbers of surface and volume atoms is 11%. Hence, for such a particle with a H atom coverage as large as one half and a charge transfer as large as 0.5 electron per H atom, the LSPR redshift would be 8 nm.

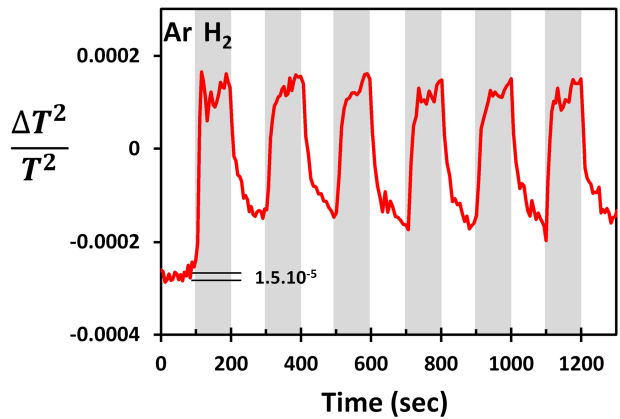

Fig.4. Change in intensity of the TA signal monitored at 613 nm during 100 s Ar / $H_2$ cycles at room temperature. The exposure to $H_2$ is due to a redshift of the LSPR. Moreover, the signal is reversible as it can be seen in the periodicity of the measurement. The average TA value has been shifted to a value close to 0, with respect to Fig. 3.b [45]

**Real-time monitoring of exposure to di-hydrogen**

In this section, the change in intensity of TA measured at 613 nm, is monitored as a function of the sample's exposure to pure Ar and $H_2$ after accurate purging of the cell and stabilisation of the signal. The gas flow was shifted between Ar and $H_2$ every 100 seconds. The result is shown in Fig. 4 for several successive cycles [45]. When shifting from Ar to $H_2$, a clear increase in TA is observed. This is equivalent to a redshift of the TA spectra (Fig.3.b). Shifting back to Ar leads to a blue shift. The signal change upon gas shifting takes about 10 seconds. This duration corresponds mainly to the time needed to purge the cell and tubes with the new gas. It takes about 500 s to completely recover from $H_2$ to Ar (not shown) which is likely due to residual $H_2$ absorbed in the cell and tube walls. The important point here is that the initial exposure to $H_2$ gives a positive increase of the TA signal of $4.0 \pm 0.2 \cdot 10^{-4}$, which corresponds to a redshift of the LSPR, equal to $\Delta\lambda = 0.50 \pm 0.02\ nm$.

The observation of this redshift upon exposing the Au NPs to $H_2$ indicates clearly that, at room temperature, $H_2$ actually adsorbs on Au NPs of lateral sizes ranging from 5 to 20 nm, even supported on a non-reducible substrate as glass. This adsorption is supposedly followed by dissociative chemisorption, as proposed by all theoretical investigations, and as shown from the previous H/D experiments [11]. As a matter of fact, when considering chemisorption on metal surfaces, the bonding between the chemisorbed entities and the metal atoms mobilises electrons, which are both donated by the adsorbate and back-donated by the metal [46,47]. Density Functional Theory (DFT) calculations have demonstrated that the dissociative chemisorption of $H_2$ on Au NPs induces significant charge redistribution within the NP. These results describe an accumulation of electrons donated both by the H atoms and, though weaker, by the Au atoms, within the Au-H molecular orbital [48]. Hence, the total number of conduction electrons involved in the Au NP's LSPR is expected to be reduced, leading to the observed redshift of the LSPR (Eq.4). The amplitude of the latter should consequently give information on the total amount of the negative charge back-donated by the Au NP to the Au-H bonds on its surface, and will be discussed below.

The possible LSPR shift caused by the Au NPs' exposure to $H_2$ has previously been reported in a couple of investigations, though the results are contradictory. Sil et al. studied the LSPR modification of a film of Au NPs ranging from 10 to 50 nm, grown by vacuum evaporation onto glass, and observed a blue shift of about 3 nm upon $H_2$ exposure [49], therefore opposite to the present observation. They explained this effect as the dissociation of $H_2$ (that they proposed to be induced by hot electrons generated by the LSPR) which led to an electron transfer from the adsorbed H to the Au NPs. On the other hand, Collins et al. did not observe any LSPR shift for Au nanorods with size 13 x 40 nm deposited on glass, at least with their margin of error equal to $\pm 0.2\ nm$ [50]. It is worth noticing that our observed redshift is of the order of the precision of the measurements in the latter reference, which explains why it could not be observed therein. Curiously, it is also six times smaller, and above all in the opposite direction, to what was measured in Ref [49] where a blue shift of 3 nm was observed. The discrepancy of our results with Ref [49] could come from the purity of the gases used in the different experiments and/or from the time during which the system was purged from air prior to the experiments in order to remove $H_2O$ and $O_2$ traces.

It can also be wondered whether our observed H chemisorption could be induced by a photochemical reaction due to the light impinging the sample, as proposed in in ref. [49]. It has been shown indeed that hot electrons induced by LSPR excitation on a metal NP can be transferred to empty electronic states of adsorbed molecules. [51] In particular, such hot electrons induce dissociation of $H_2$ molecules on Au NPs at room temperature, either supported on $SiO_2$, $Al_2O_3$ or $TiO_2$ [11,52]. However, in these experiments, laser light of intensity larger than 0.25 W/cm$^2$ was necessary for this photochemical effect to take place on top of the observed smaller spontaneous dissociation. In the present experiments, the illumination intensity was about 0.01 W/cm$^2$, which was too low to induce any additional dissociation. This was verified, by performing with another sample several $H_2$ / Ar cycles with illumination intensities varying from $10^{-2}$ to $2.10^{-4}$ W/cm$^2$, as shown in Fig.5. The signal amplitude during the gas cycles is identical, regardless of the illumination energy. The only noticeable difference is the expected lower signal/noise ratio for smaller light intensities. This demonstrates that the chemisorption and dissociation of $H_2$ by the Au NPs observed in the present experiment is not induced by the incident light.

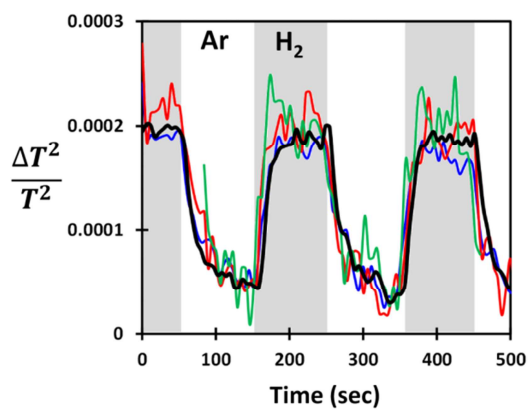

Fig.5. H$_2$ / Ar cycles for a Au sample with various intensities of impinging light. Black:100% intensity (about 0.01 W/cm$^2$), blue: 25%, red: 7%, green: 2%. Although the signal to noise ratio decreases when the light intensity decreases, it is clear that the amplitude of the shift is not related to the light intensity, meaning that the process is not photo-activated.

**Adsorption, diffusion and desorption of hydrogen**

Several samples with different quantities of Au have been investigated. For every sample, the change of the TA signal has been measured and compared to the average size of the NPs, determined by SEM for their lateral size and AFM for their height. Hence, by using the shape of the measured TA spectrum, and by knowing the average volume of the NPs of each sample and therefore the number N of conduction electrons, the corresponding total charge transfer ΔN per particle could be calculated, using eq4. The evolution of ΔN with the average diameter of the NPs is drawn in Fig 6.a.

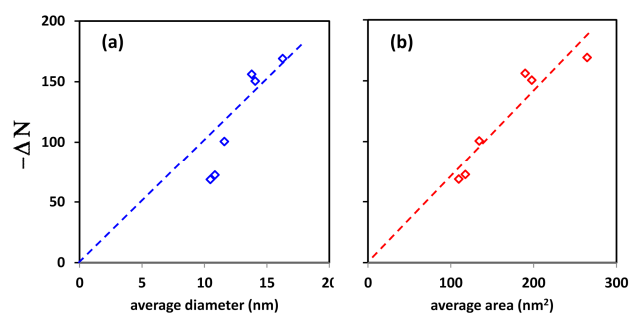

Fig. 6. Dependence of -ΔN on the average diameter of the NPs (a) and on the average area of the NPs (b). The dashed lines corresponding to proportional relationship are drawn as guides for eyes. These results clearly show that ΔN is proportional to the average surface area of the NPs and not to the average diameter. This dependence can be explained by considering the migration of the adsorbed H atoms towards the Au facets, after dissociation at the edges.

It shows that ΔN is not proportional to the NPs' size, as it could have been expected from the above discussion which indicates that the H$_2$ molecules are dissociated on the low-coordinated Au atoms located on the edges, which amount is proportional to the NP size. Quite the opposite, Fig.6.b shows that the charge transfer ΔN is proportional to the area of the NPs. This, at first sight surprising, dependence of ΔN on the NPs' area can be explained by considering the migration of the adsorbed H atoms towards the Au facets after dissociation at the edges. As a matter of fact, the surface diffusion barrier of H has been recently calculated by DFT on the dense surfaces of Au, by considering hopping between different sites. It has been found to be equal to 0.15 eV for the (111) surface, and is of the same order or even smaller on the (100) one [53]. This relatively small energy leads to a very fast diffusion on the Au surfaces [54]. This should also insure a uniform H coverage on each kind of facets. On the other hand, the binding energy for two H atoms on the most stable fcc position of Au(111) has been calculated by DFT [53] and is $2\,E_{(111)} = -4.36\,eV$, higher than the formation energy of H$_2$ ($-4.57\,eV$ calculated, against $-4.52\,eV$ experimental), which indicates that the H atoms diffusing on this surface are not stable and recombine quickly into H$_2$. The calculated binding energy of two H atoms on bridge position of the (100) facet is equal to $2\,E_{(100)} = -4.54\,eV$, hence almost the same as the calculated energy of formation of the molecule. Moreover, the difference of 0.03 eV is close to the thermal energy at room temperature. It is therefore reasonable to anticipate that H atoms, issued from H$_2$ dissociation on NP edges and migrating on the Au NPs (100) facets, could stay a sufficient time before recombining into molecules, so that the occupation in H of the Au sites on these facets is non-negligible, giving rise to charge transfer and therefore to the observed redshift of the LSPR. Lastly, the equilibrium equation between the coverages of H atoms on the (100), $\theta_{(100)}$ and on (111) facets, $\theta_{(111)}$, can be written:

$$\theta_{(100)}\,e^{\frac{E_{(100)}}{k_B T}} = \theta_{(111)}\,e^{\frac{E_{(111)}}{k_B T}} \qquad (5)$$

which gives a ratio of $\frac{\theta_{(111)}}{\theta_{(100)}} = 0.03$ with the above binding energies, and of 0.015 with adsorption energies calculated in Ref [4]. As the ratio between (111) and (100) facets on an Au NP with the Wulff equilibrium shape is about 3, it indicates that the number of sites occupied by H on the (111) facet remains quite smaller than that occupied on the (100) facets. Hence only the latter are considered in the following analysis. Other facets like (110) and (211) have larger surface energies and are not present on Au NPs, as well known by transmission electron microscopy studies. [55,56]

In order to investigate the kinetic mechanisms in play, experiments have been performed with varying flow of incident H$_2$ molecules. Following the above reasoning, the amount of H on the (111) facets should not be modified and stay very small, whereas it should increase on the (100) facets with increasing flow, leading to a larger charge transfer, thus a larger redshift. The flow of incident H$_2$ has been varied by changing the partial H$_2$ pressure *p* in the Ar/H$_2$ mixture within the cell, maintained at atmospheric pressure, from 67% to 1%. Figure 7 gives the intensity of the measured redshift for a given sample, which strongly decreases for decreasing H$_2$ partial pressure.

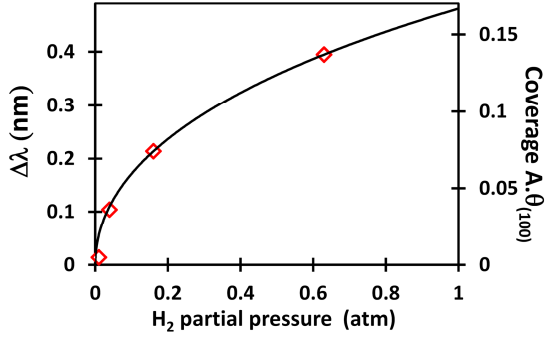

Fig 7. Red lozenges: Experimental redshift of the LSPR upon H$_2$ adsorption (left scale); black continuous line: H coverage $\theta$ of the NP facets, obtained from Eq.12 (right scale). The height of the lozenges corresponds to the error bar.

The kinetics of dissociative adsorption and associative desorption of molecules has been long-time investigated.[55,56] In the present case, one can consider that the adsorption occurs on the edges of the NPs, where the molecules dissociate, with an H coverage $\theta_{edge}$. Then the H atoms either desorb back to the gas phase or can migrate on the facets. On the (100) facets, they can either stay a finite time and diffuse with coverage $\theta_{(100)}$, recombine and desorb back to the gas phase, or migrate back to the edges. The diffusion rate on the (100) facets being very fast, the coverage $\theta_{(100)}$ can be taken as homogeneous. Applying the stationary state approximation for the edges and for the facets, the corresponding kinetic equations at equilibrium can be written:

$$\frac{d\theta_{edge}}{dt} = 0 = r_a(\theta_{edge}) - r'_d(\theta_{edge}) - r_{mig}(\theta_{(100)}, \theta_{edge}) + r'_{dmig}(\theta_{(100)}, \theta_{edge}) \quad (6)$$

$$\frac{d\theta_{(100)}}{dt} = 0 = r_{mig}(\theta_{(100)}, \theta_{edge}) - r'_{mig}(\theta_{(100)}, \theta_{edge}) - r_d(\theta_{(100)}) \quad (7)$$

$r_a(\theta_{edge})$ is the adsorption rate at the edges, $r'_d(\theta_{edge})$ and $r_d(\theta_{(100)})$ are the desorption rates from the edges and the (100) facets, $r_{mig}(\theta_{(100)}, \theta_{edge})$ and $r'_{mig}(\theta_{(100)}, \theta_{edge})$ are the migration rates from the edges to the facets and from the facets to the edges, respectively. The desorption and adsorption rates, for the case of non-interacting adsorbed atoms and without considering precursor state, for simplicity, have the following second-order Langmuirian expressions, depending on the coverages: [56,57]

$$r_a(\theta_{edge}) = k_a(T)\, \Phi\, (1 - \theta_{edge})^2 \quad (8)$$

$$r_d'(\theta_{edge}) = k_d'(T)\, \theta_{edge}^2 \quad (9)$$

$$r_d(\theta_{(100)}) = k_d(T)\, \theta_{(100)}^2 \quad (10)$$

where $k_a(T)$, $k_d'(T)$ and $k_d(T)$ are the rate coefficients of dissociative adsorption on the edges and associative desorption from the edges and facets, respectively, which are dependent on the temperature T. $\Phi$ is the flow of impinging molecules from the gas phase, given by $p/(2\pi m k_B T)^{1/2}$, where p is the pressure. Calculating the difference between Eqs. 6 and 7 leads to suppress the unknown migration rates, and yields the simple expression:

$$k_a(T)\,\Phi\,(1-\theta_{edge})^2 - k'_d(T)\,\theta_{edge}{}^2 - k_d(T)\,\theta_{(100)}{}^2 = 0 \quad (11)$$

In order to determine the ratio between $\theta_{(100)}$ and $\theta_{edge}$, the adsorption energies on the edge and on the facet should be calculated within the same theoretical model. To our knowledge, such comparison has not been performed. Consequently, the ratio $\frac{\theta_{edge}}{\theta_{(100)}}$ was taken in the following as a constant A, which is likely close to or larger than 1. After renormalising the flux by using: $F = k_a(T)/k'_d(T)\,\Phi$, the coverage $\theta_{(100)}$ can be written:

$$A\,\theta_{(100)} = \frac{F^{1/2}}{F^{1/2} + \left(1 + \frac{k_d}{A^2 k'_d}\right)^{1/2}} \quad (12)$$

This expression, with only one parameters ($\frac{k_d}{A^2 k'_d}$) allows one to reproduce the experimental points of Fig.7 very well. Within this approach, the H coverage on the (100) facets can be obtained (with the factor A), and its value, read in Fig.7 on the right hand side scale is $A\,\theta_{(100)} = 0.16 \pm 0.05$ at atmospheric pressure. This large uncertainty is due to the fact that the evolution of the experimental points in Fig.6 can be reproduced correctly, within the error bars, for a range of values of $\frac{k_d}{A^2 k'_d}$, each one modifying the right scale of Fig.6.

The effect of temperature on the intensity of the LSPR shift is also informative and has been investigated. Fig 8 shows that the LSPR shift during the Ar / $H_2$ cycles strongly decreases with increasing temperature. This can be explained by a reduced time for the recombination of H atoms chemisorbed on the Au NP at high temperatures, leading to a faster desorption of $H_2$ molecules, and therefore a smaller H population on the Au NPs. This reduced time can be related to a faster diffusion on the facets and to an easier overcoming of the possible activation barrier for $H_2$ recombination. However, a simple behaviour involving only one activation energy, following either an Arrhenius law or Equation 12, could not be determined from the present temperature dependence of the LSPR shift. Nonetheless, such increase in the $H_2$ recombination explains both the decrease of the LSPR shift measured herein and, on the opposite, the increase of the HD formation with higher temperatures during H/D exchange experiment observed by Mukherjee et al, which actually corresponds to an increase of the turn-over-frequency, related to the increase of both the dissociation rate and recombination rate. [11].

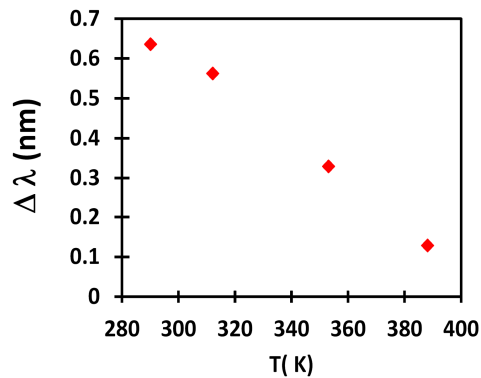

Fig.8. LSPR shift measured as a function of the sample's temperature. There is a clear decrease of the shift when the temperature rises, explained by a faster recombination of the H atoms.

**Charge transfer between Au and H**

The charge transfer $\Delta N$ induced by H adsorption can now be calculated with Eq.4. This gives, for the results presented in Fig.4, $\Delta N/N = -1.5 \times 10^{-3}$. In order to relate this value to the charge transfer, the number of adsorbed H atoms in permanent regime has to be estimated. Considering that only the (100) facets are covered by H, the NP shape is needed. Au NPs prepared by vacuum deposition at RT have been observed during epitaxial growth on MgO surface by transmission electron microscopy, and display regular (111) and (100) facets with few defects [55]. Although Au NPs do not grow with such an epitaxy on a glass substrate at RT, they are expected to also display well-defined facets. For simplicity, if considering that all particles have the shape of the upper part of a truncated octahedron (which is the equilibrium shape of Au NPs), either with a (111) or a (100) facet in contact with the substrate, the average area of (100) facets for NPs of lateral size equal to 12 nm and height equal to 4.5 nm is about 50 nm$^2$, which contains 600 atoms (of which 150 on the edges). The average volume of these NPs is about 400 nm$^3$, which contains 24 000 Au atoms. Exposed to pure H$_2$, the negative charge transfer is therefore $\Delta N = -35$ yielding an average charge transfer equal to $-0.06\ e$ per (100) surface Au atom. However, as the ratio A between $\theta_{edge}$ and $\theta_{(100)}$ is unknown, it is not easy to know the actual H coverage on the (100) facets. Due to the fact that the adsorption sites on the edges [4] and on the (100) facets [53] are similar bridge sites, we can consider that this ratio A is close to 1. Consequently, Fig. 7 indicates that the coverage under atmospheric pressure of H$_2$ is about $\theta_{(100)} = 0.16 \pm 0.05$, which would correspond to $96 \pm 30$ H atoms on the (100) facets. Therefore, one can estimate that an amount $0.2 \pm 0.05$ electron is back-donated from Au to each Au-H bond. Comparing this result with theoretical results is not straightforward. A few investigations of the charge transfer between H and Au NPs have been performed, but they are somehow contradictory. For instance, Kuang et al. found a small negative charge of about $-0.1\ e$ in the Au$_n$ part of Au$_n$H$_2$ clusters with dissociated H$_2$, [59] while Lyalin et al. found no or smaller charge transfer for Au clusters, either isolated or supported [60]. On the contrary, Zhao et al calculated a positive charge equal to 0.25 and 0.3 e in the Au parts of Au$_4$H and Au$_5$H clusters [61]. Similarly, Libisch et al calculated an electron transfer from Au to an adsorbed H$_2$ molecule, corresponding to a partial occupation of the σ* antibonding orbital of H$_2$, reaching the value $-0.8\ e$ at the top of the barrier, which favoured the dissociation [62]. Finally, Hu et al, for Au$_{85}$ clusters, showed that H adsorption leads to a significant reorganisation of the charge distribution, with a depletion of electrons, both in the Au atoms in the vicinity of the adsorbed H and on the H atom, explained by the back-donation and the donation of electrons, from Au and H atoms, respectively [48]. Consequently, our results showing a decrease of the electron population in the Au NP upon H adsorption, appear in line with these three latter theoretical results.

# Conclusion

By investigating the shift of the LSPR Au NPs of size around 10 nm upon exposure to di-hydrogen, we confirmed the actual H$_2$ chemisorption on Au. The non-reactive glass substrate used here indicates that this chemisorption occurs most likely directly on the NPs. The most important information is the observation of an electron transfer from Au to H, which decreases the density of charge in the Au NPs and, consequently, yields a redshift of the LSPR of a fraction of nm. Such small redshift could be determined accurately by use of a differential optical technique, applied to anisotropic samples elaborated in an easy way. Moreover, the proportionality of the intensity of the redshift with the area of NPs with different sizes show that the dissociated H atoms migrate on the facets of the NPs, most likely the (100) facets, where they eventually recombine and desorb back to the gas phase. At atmospheric pressure of $H_2$, the average negative charge transfer from every surface Au atoms to H is about - 0.06 e, and the charge back-bonded localised on every Au-H bonding is estimated to be around - 0.2 e. This deeper understanding of the $H_2$ chemisorption mechanism onto Au NPs is hoped to help improving the elaboration and use of catalytic Au NPs for hydrogenation reactions.

## Conflicts of interest

There are no conflicts of interest to declare.

## Acknowledgements

The authors are grateful to Sébastien Royer for his technical assistance and Dominique Demaille for a help concerning the use of the SEM apparatus. Geoffroy Prévot is thanked for illuminating discussions, in particular concerning the kinetics equations, and for critical reading of the manuscript.

## Notes and references

**SUPPLEMENTARY INFORMATION**

**Calculation of the optical properties of flat supported Au nanoparticles.**

Let us consider a Au NP with a flat ellipsoidal shape, close to the spheroid but with a slight anisotropy, with half-axes corresponding to the average values determined by AFM and MEB: $a = 5.5\ nm$, $b = 6.5\ nm$ and $c = 2.25\ nm$ (Fig S.1). Here we neglect the interaction between the NPs, which is another way to take into account an overall anisotropy of the substrate. It is known that the interaction with the substrate yields a red shift of the LSPR of a NP. However, taking into account rigorously such effect is not simple and, to our knowledge, treatment has been developed for spheroids only and cannot be applied for ellipsoidal particles.[1,2] Nevertheless, a simple way to introduce approximately the effect of the glass substrate is to consider, an intermediate dielectric function $\varepsilon_{emb}$ between those of air ($\varepsilon_{air} = 1$) and of the substrate ($\varepsilon_{glass} = 2.25$), for the embedding medium of the NPs.[3] Moreover, an increase of the damping of the free electrons of Au, due to the small size of the NPs, has been considered in the calculation, which damps out and enlarges the LSPR. With the above values, the transmittance square is given in Fig.S2, for a mass thickness equal to 2 nm. The resonance is located around 600 nm along the x direction (the short size of the ellipsoid parallel to the surface) and shifted to 700 nm along the y direction (the long size), as illustrated in Figure S1. This is in agreement with the experiment. Here, the aim of the calculation was not to perfectly reproduce the experiment but to demonstrate that the measured sizes of the NPs could explain the observed positions of the LSPR. For a better agreement, several leads could be followed: (i) taking into account the distribution of size and of shape of the particles would enlarge the resonances; (ii) taking into account the electromagnetic interaction between the particles (as the apparent chains of particles observed in Fig.2) would increase the anisotropy and redshift the LSPRs. Finally, treating more rigorously the interaction with the substrate could be done, either by using approaches developed for spheroids and considering that the present NPs have shapes close to spheroids, or by using finite element methods such as Discrete Dipole Approximation, although in this model too, the interaction with the substrate is not easily taken into account.

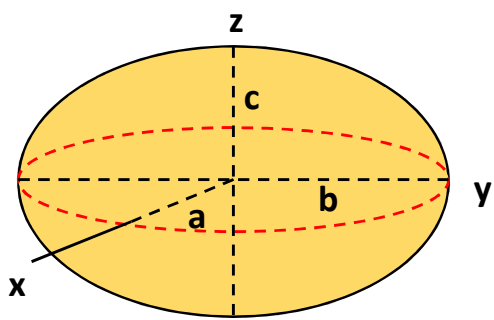

**Fig. S1.** Ellipsoid diagram representing an anisotropic shape. The x (short) and y (long) axes show the directions of the polarised light in relation to the particle.

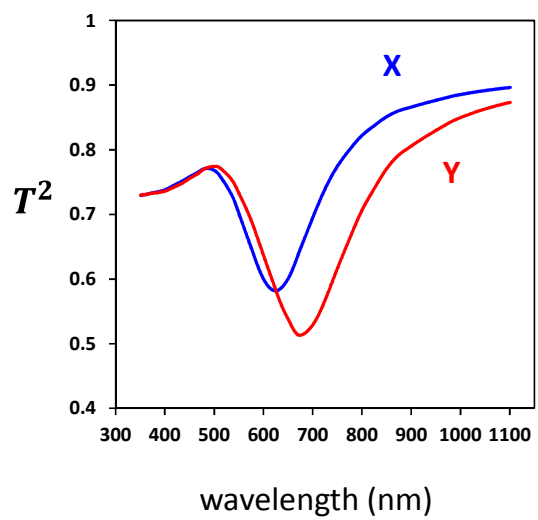

**Fig. S2.** Transmittance squared of an ensemble of Au ellipsoids on a glass substrate, with mass thickness of 2 nm

**TAS results for different samples.**

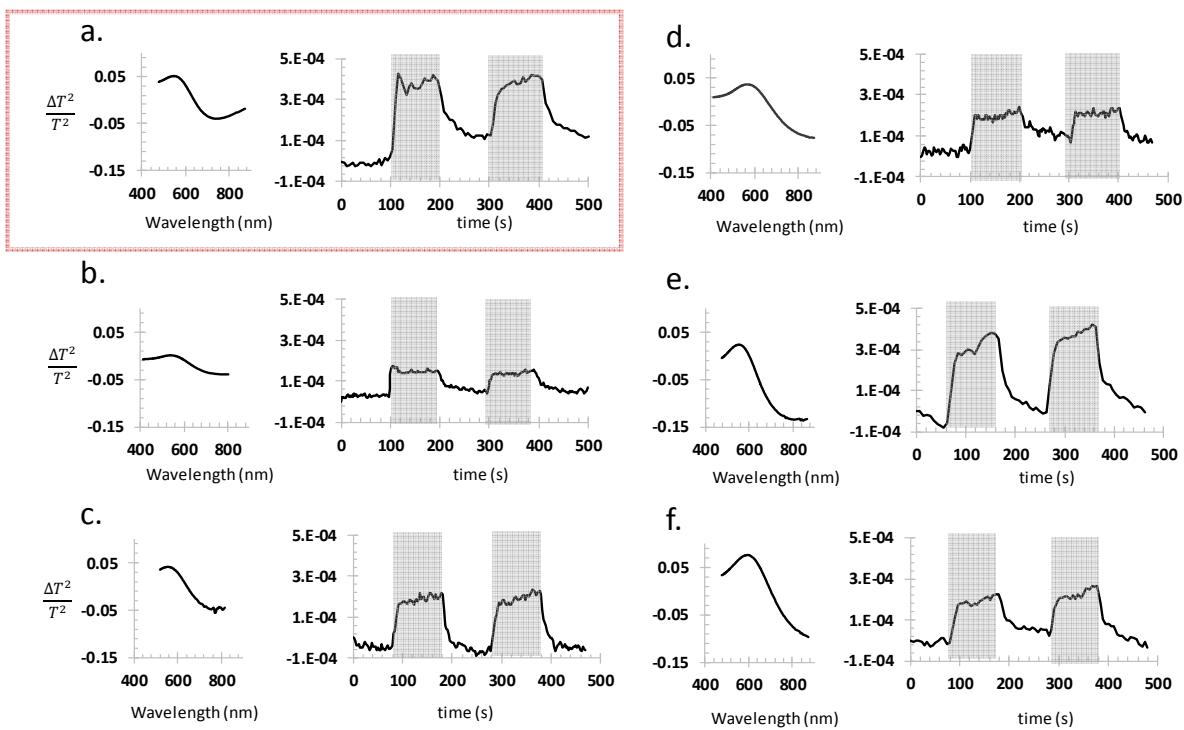

**Fig S3:** Graphs for different samples. Each pair is for a given sample of Au nanoparticles of average diameter: a. 10.5nm b. 10.8nm c. 11.6nm d. 13.7nm e. 14.1nm f. 16.3nm. The figure on the left hand side of a pair correspond to a TAS spectrum of the sample and its corresponding pair is a real time measurement of the change in intensity of the TA signal when exposing the sample to Ar and $H_2$ in 100s cycles. As each sample qualitatively displays the same behaviour, detailed results are only given for sample (a) in the paper.